\documentclass[article, superscriptaddress, floatfix]{revtex4-1}
\usepackage[english]{babel}
\usepackage{graphicx}
\usepackage{amsmath}
\usepackage{amssymb}
\usepackage{textcomp}
\usepackage{xcolor}
\usepackage{hyperref}
\setcitestyle{super}

\begin{document}

\title{Crystallography companion agent for high-throughput materials discovery}

\author{Phillip M. Maffettone}
\email[]{pmaffetto@bnl.gov}
\affiliation{National Synchrotron Light Source II, Brookhaven National Laboratory, Upton, New York 11973, USA}
\affiliation{Department of Chemistry and Materials Innovation Factory, University of Liverpool, Crown Street, Liverpool L69 7ZD, U.K.}
\author{Lars Banko}
\affiliation{Institute for Materials, Faculty of Mechanical Engineering, Ruhr University Bochum, 44801 Bochum, Germany}
\author{Peng Cui}
\affiliation{Department of Chemistry and Materials Innovation Factory, University of Liverpool, Crown Street, Liverpool L69 7ZD, U.K.}
\author{Yury Lysogorskiy}
\affiliation{Interdisciplinary Centre for Advanced Materials Simulation (ICAMS), Ruhr University, 44801 Bochum, Germany}
\author{Marc A. Little}
\affiliation{Department of Chemistry and Materials Innovation Factory, University of Liverpool, Crown Street, Liverpool L69 7ZD, U.K.}
\author{Daniel Olds}
\affiliation{National Synchrotron Light Source II, Brookhaven National Laboratory, Upton, New York 11973, USA}
\author{Alfred Ludwig}
\affiliation{Institute for Materials, Faculty of Mechanical Engineering, Ruhr University Bochum, 44801 Bochum, Germany}
\author{Andrew I. Cooper}
\email[]{aicooper@liverpool.ac.uk}
\affiliation{Department of Chemistry and Materials Innovation Factory, University of Liverpool, Crown Street, Liverpool L69 7ZD, U.K.}

\date{\today}

\begin{abstract}
The discovery of new structural and functional materials is driven by phase identification, often using X-ray diffraction (XRD). Automation has accelerated the rate of XRD measurements, greatly outpacing XRD analysis techniques that remain manual, time-consuming, error-prone, and impossible to scale. With the advent of autonomous robotic scientists or self-driving labs, contemporary techniques prohibit the integration of XRD. Here, we describe a computer program for the autonomous characterization of XRD data, driven by artificial intelligence (AI), for the discovery of new materials. Starting from structural databases, we train an ensemble model using a physically accurate synthetic dataset, which output probabilistic classifications---rather than absolutes---to overcome the overconfidence in traditional neural networks. This AI agent behaves as a companion to the researcher, improving accuracy and offering significant time savings. It was demonstrated on a diverse set of organic and inorganic materials characterization challenges. This innovation is directly applicable to inverse design approaches, robotic discovery systems, and can be immediately considered for other forms of characterization such as spectroscopy and the pair distribution function.
\end{abstract}

{
\let\clearpage\relax
\maketitle
}
\section{Introduction}
Phase identification using X-ray diffraction (XRD) is a linchpin in the discovery of materials for diverse applications including batteries, catalysis, and pharmaceuticals. Automation has accelerated the rate of XRD measurements, greatly outpacing XRD analysis techniques that remain for the most part manual, time consuming, error prone, and impossible to scale. This prevents the integration of this essential technique with autonomous robotic searches or self-driving labs \cite{Granda_2018,  MacLeod_2020, Burger_2020}. Artificial intelligence (AI) can assist in the classification of XRD patterns\cite{Iwasaki_2017, Stanev_2018, Xiong_2017, Long_2009, Takeuchi_2005, Oviedo_2019, Lee_2020, Ziletti_2018, Aguiar_2019, Chen_2020}, but widespread adoption is challenging due to limited reproducibility beyond specific materials systems\cite{Stanev_2018,Oviedo_2019, Park_2017,Chen_2020}. Here we report an AI approach for the autonomous phase identification of diffraction patterns that is accurate across both organic and inorganic materials systems. We created a crystallography companion agent (XCA)—an algorithm-powered tool to collaborate with the researcher—that achieves expert accuracy in real-time with the measurements, using both experimental and predicted crystal structures as inputs. XCA overcomes the overconfidence of traditional neural networks through a probabilistic strategy that can incorporate multimodal analysis. This is accomplished without manual human-labelled data and is robust against many sources of complexity in diffraction. It is also extendable to other forms of characterization that can be accurately simulated, such as spectroscopy. This development complements recent advancements in automation\cite{King_2011, Li_2015, Dragone_2017, Granda_2018, Buenconsejo_2014, Langner_2020, MacLeod_2020, Steiner_2019, Bedard_2018}  and autonomous experimentation\cite{Granda_2018,  MacLeod_2020, Burger_2020}, thus enabling a key step in the accelerated materials analysis.
\par

\begin{figure}
  \centering
  \includegraphics{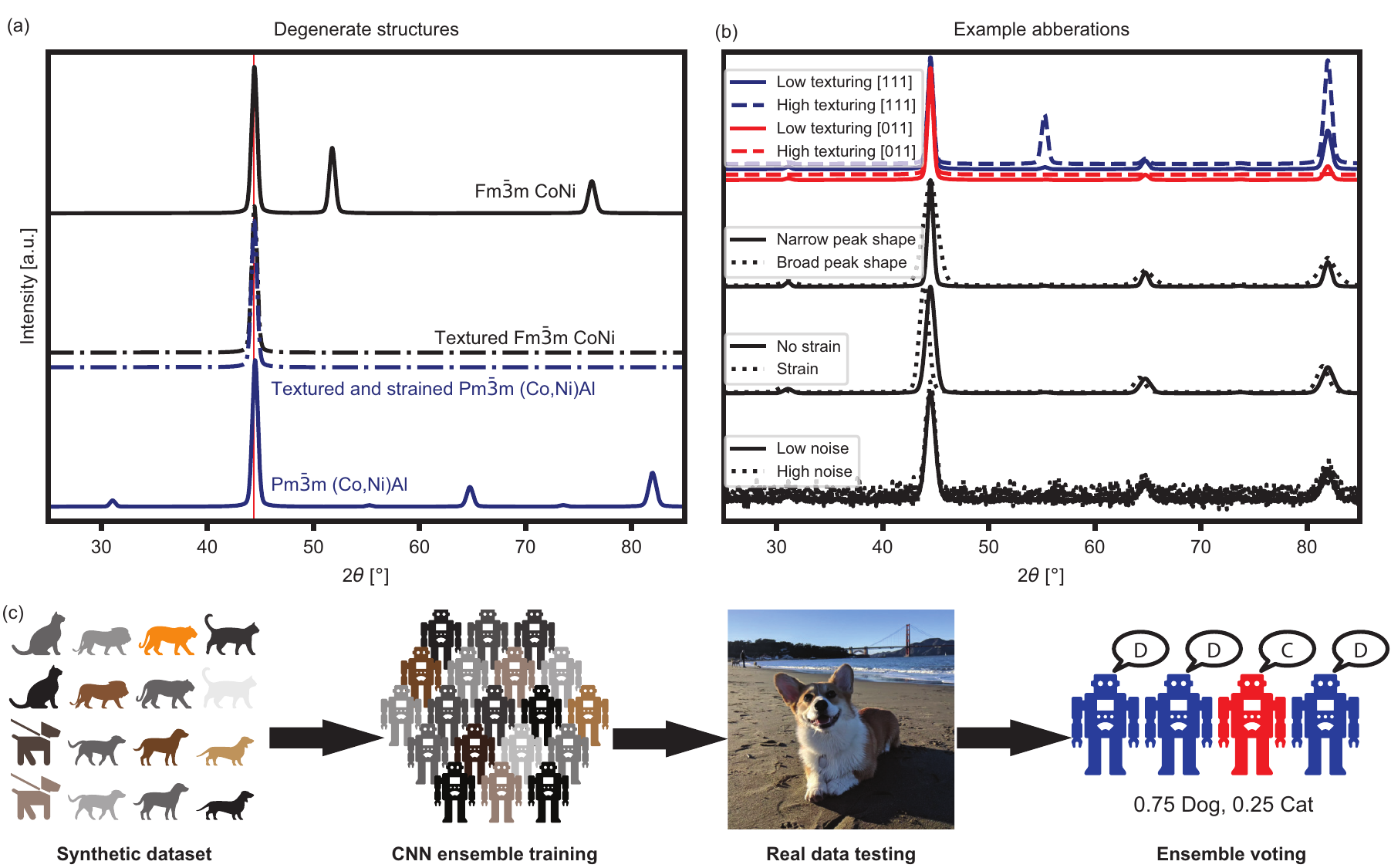}
  \caption{\textbf{Experimental XRD complexity and training an ensemble using synthetic data.}
    \textbf{a,} Different crystal phases can create identical XRD patterns under conditions that are common in thin-film experiments.
    \textbf{b,} There are many causes of aberration in an XRD pattern, such as intensity changes from preferred orientation, peak shifting from lattice strain or solid solutions, peak broadening, and background, as illustrated here for the Pm$\overline{3}$m phase of (Ni,Co)Al.
    \textbf{c,} The crystallography companion agent (XCA) statistically solves the problem of simultaneous experimental complexity and data scarcity by automatically building a synthetic dataset and training an ensemble of learners from this data. The dataset covers the scope of variation in XRD patterns and the ensemble model outputs an existence probability of each phase when tested against real data. This protocol is analogous to training a cat-vs-dog classifier on artistic sketches of the animals and testing on photographs. Unlike the sketch analogy, this training approach is possible for XRD because of the speed and accuracy of the simulations.
  }
  \label{fig1}
\end{figure}

Even with the help of dedicated software, the analysis of XRD patterns to determine unknown phases is challenging, error prone, and time consuming. Multiple sources of aberration affect experimental XRD patterns, affecting peak shapes, positions, and intensities (Fig.~\ref{fig1}a, b), leading to degenerate patterns. This is compounded by the problem of homometrics\cite{Patterson_1939} (Fig.~\ref{fig1}a), where multiple unique structures can equivalently explain an XRD pattern. Thus, a given crystal phase can correspond to many unique XRD patterns, and an XRD pattern can correspond to multiple unique phases. Some modern XRD instruments can measure hundreds, if not thousands, of patterns per hour, while the analysis of a single novel pattern can take hours or even days, and this introduces a significant bottleneck to discovery. Furthermore, both inorganic\cite{Collins_2017} and organic\cite{Pulido_2017} crystal structure prediction (CSP) are increasingly useful tools for the prediction of stable crystalline materials with useful functional properties, but here the XRD data problem is further compounded by the need to match large numbers (1000’s) of predicted patterns with experimental data that may not match the theoretical predictions precisely, either because of limitations in the predictions, the measurements, or often both.
\par

Emerging big-data applications in materials science show promise as tools for XRD analysis. Pattern matching is commonly used to compare XRD patterns to reference structures\cite{Ivanisevic_2005, Huang_1981, Gregoire_2011, Stein_2017}, either by hand or with peak matching algorithms, sometimes incorporating additional constraints; for example, the Gibbs phase rule\cite{Xiong_2017, Ermon_2015, Xue_2017, Kusne_2015, Suram_2017}.
These methods do not account for common effects such as preferred orientation, peak shifting, or phase mixtures. Unsupervised methods attempt to statistically segregate experimental patterns for further analysis\cite{Iwasaki_2017, Stanev_2018, Xiong_2017, Long_2009, Takeuchi_2005}.
These methods are useful when there are no data on expected phases and they can be combined with traditional forms of structure solution. However, unsupervised methods are highly susceptible to experimental variation, which can lead to an overestimation of the number of distinct phases\cite{Stanev_2018}. More recently, semi-supervised deep learning has been shown effective for inorganic XRD\cite{Oviedo_2019, Ziletti_2018, Park_2017}, convergent beam electron diffraction12 and electron backscatter diffraction\cite{Kaufmann_2020}.  Many of these supervised methods are reliant on large proprietary datasets \cite{Kaufmann_2020, Park_2017, Lee_2020}, suffer from combinatorial explosion\cite{Lee_2020}, and remain over-confident in their predictions, offering no measure of uncertainty\cite{Blundell_2015}.
These models make use of physical knowledge and are trained successfully on partially\cite{Oviedo_2019, Wang_2020} or completely simulated datasets\cite{Lee_2020}.
To date, these machine learning methods have only been demonstrated as accurate in specific domains where there are available test cases; that is, these approaches are only predictive for certain classes of materials, frequently inorganic oxides\cite{Stanev_2018, Park_2017, Oviedo_2019, Lee_2020}.
\par

Our objective was to build a computer program to assist in phase identifications from large experimental XRD datasets of diverse materials systems. Such a tool requires rapid predictions, a high degree of automation, and accuracy on par with an expert crystallographer.
If a realistic dataset can be synthesized, as is the case for diffraction, then supervised learning can be implemented without manually labelling data. A synthetic dataset needs to capture the underlying physics of the measurement and the diversity of patterns caused by experimental non-idealities (Fig.~\ref{fig1}b). This approach is analogous to attempting to train a classification model to recognize photographs based on hand-drawn sketches; it is not feasible for sketches because they are insufficiently realistic and cannot be produced \emph{en masse}, but such an approach is possible in the physical sciences (Fig.~\ref{fig1}c). These datasets should be embedded in a model that is accurate, but not overconfident, and capable of integrating other prior information, such as composition.
The entire protocol should also be easy to use by researchers who are not specialists in machine learning.
\par

To address this challenge, we developed an autonomous crystallography companion agent (XCA) that learns from fully synthetic data and can predict phases from XRD patterns in real-time. The parameters that govern the dataset construction exploit the same prior knowledge of a researcher that is required for experiment preparation; for example, sample composition and diffraction instrument parameters. Under this paradigm, the scientist remains sovereign over the research while the companion agent autonomously prepares analyses under the researcher’s direction.

\section*{Results}

\begin{figure}
    \centering
    \includegraphics{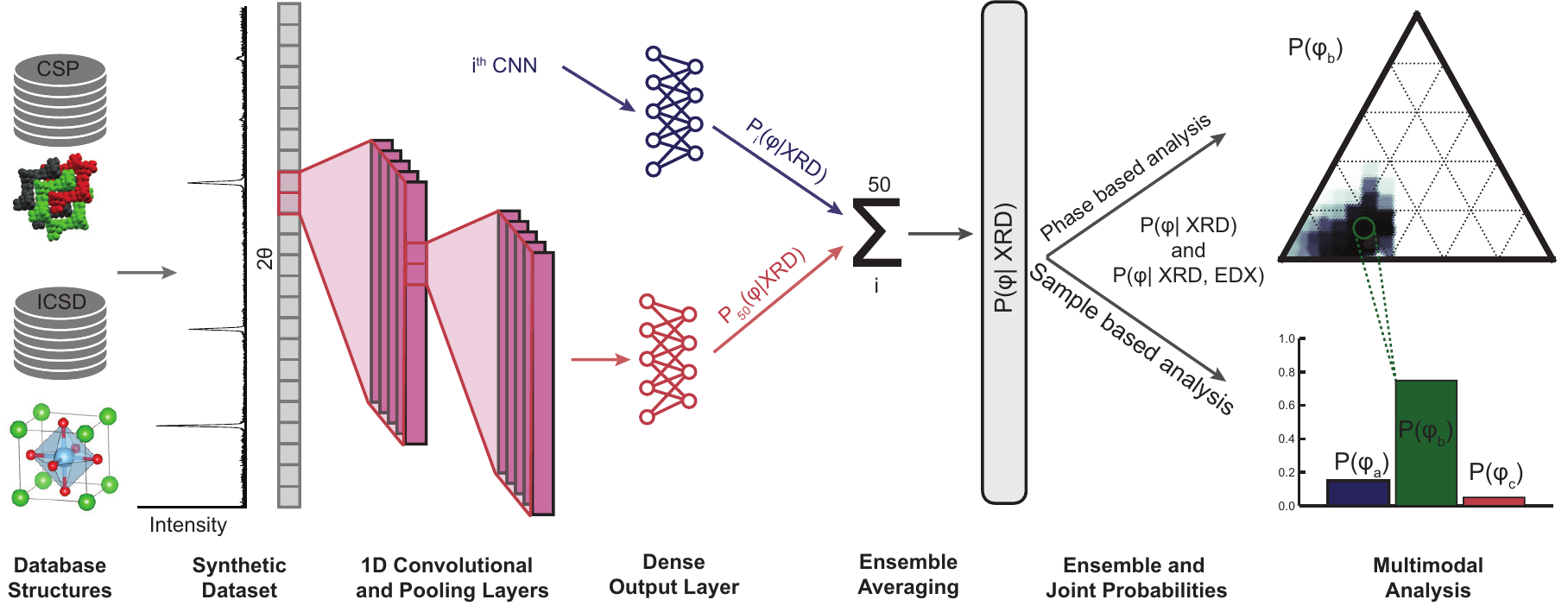}
    \caption{\textbf{Schematic of the crystallography companion agent (XCA).}
      Using only structures from databases and experimental information as inputs, XCA builds a realistic dataset of XRD patterns, and trains an ensemble of convolutional neural networks (CNNs). A single CNN learner is composed of alternating convolutional and pooling layers, followed by a single dense layer. A convolutional layer extracts features from its preceding layer, using filters learned during training, to form feature maps. The pooling down-samples the feature maps to exploit locality. The final feature maps are combined and fed to a dense layer—a simple type of classification model—that computes an output probability that the input diffraction pattern belongs to a given phase. The model averages the output of many learners to solve the problem of overconfidence in an individual CNN. Once trained, XCA will take an XRD pattern and output a probability, $P$, over all proposed phases in a few milliseconds. This $P(\phi|\mathrm{XRD})$ can then be conflated with other probability distributions; for example, those based on measured composition from EDX or, potentially, relative lattice energies from crystal structure prediction. 
      }
    \label{fig2}
  \end{figure}

XCA generates a synthetic dataset from phases within databases that encompasses the range of experimental variation for a given materials system and experimental set-up (Fig.~\ref{fig1}b).
These data are then used to train an ensemble of fifty convolutional neural networks (CNNs) that output a probability distribution over the input phases, $P(\phi|\mathrm{XRD})$ (Fig.~\ref{fig2}).
This can be conflated with independent distributions from calculated phase stability, thermodynamic constraints (such as Gibbs phase rule or phase diagram connectivity constraints), or multimodal analysis, exemplified here using energy dispersive X-ray spectroscopy (EDX).
Building the dataset and model takes a few hours on a dedicated desktop. Analysis can then be conducted in realtime for each sample, or across each phase for a full experimental dataset.
\par

Analysis proceeds on a phase-by-phase basis by mapping the likelihood of a phase across compositional space, or a sample-by-sample basis by exploring the probability of all phases in a given sample. This process is fully automated for finding pure phases of interest. The output offers a qualitative measure of phase mixing, and it is therefore suited for complete combinatorial phase mapping of multinary materials libraries. Starting from structural databases, XCA learns diffraction asynchronously with high-throughput experiments, and thus provides real-time, probabilistic analysis for the researcher. This creation of an AI tool that translates XRD measurements to probabilistic phase mapsis a necessary innovation to enable autonomous materials experimentation.

\par
\begin{figure}
  \centering
  \includegraphics{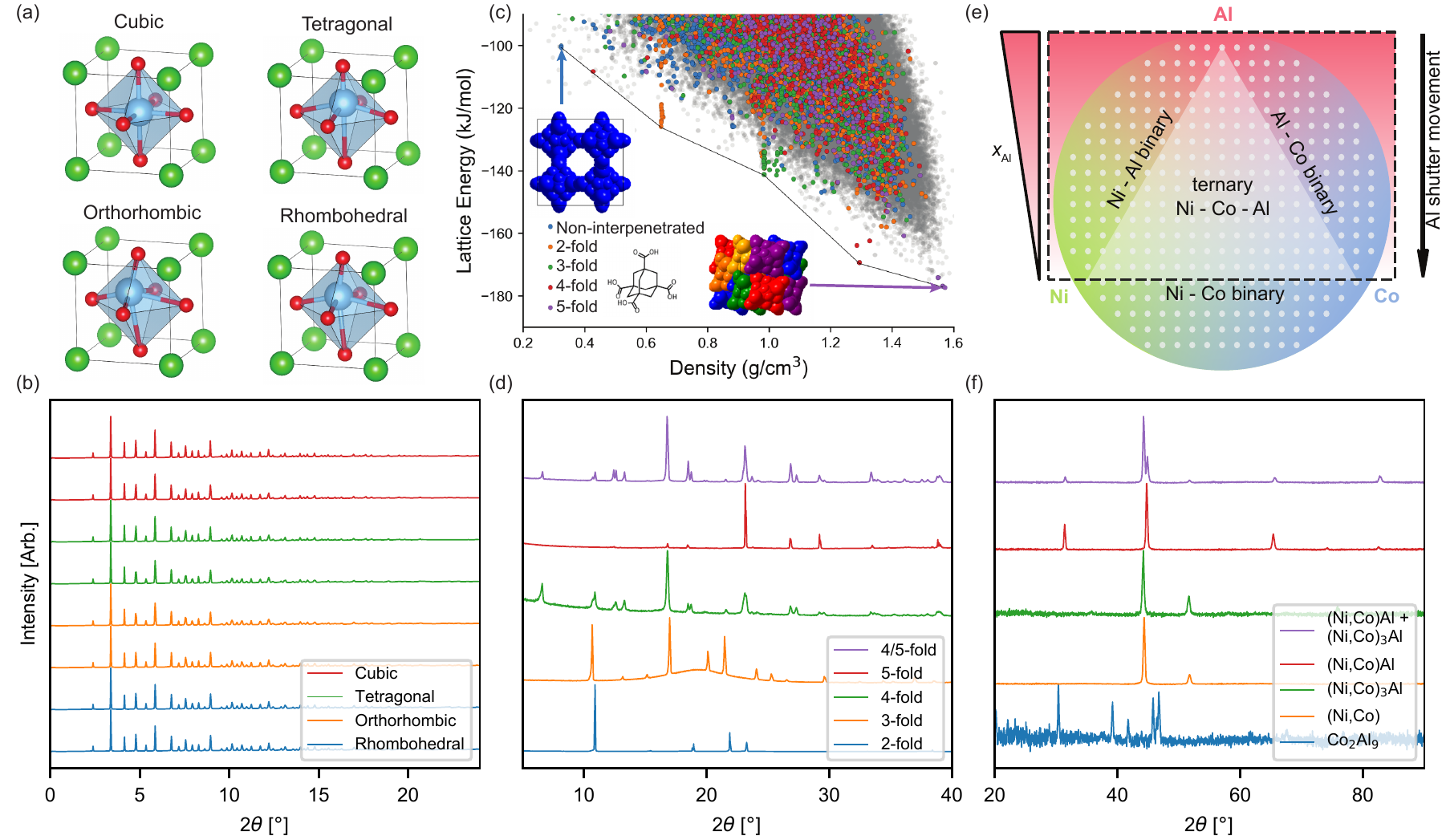}
  \caption{\textbf{Testing XCA against three different inorganic and organic materials challenges.}
    \textbf{a,} Phase transitions in BaTiO$_3$ involve symmetry breaking from a Ti translation.
    \textbf{d,} These phases are hard to distinguish by XRD.
    \textbf{b,} Crystal structure prediction for ADTA\cite{Wang_2020} identified five low-energy phases (end members shown here) with increasing degrees of interpenetration that were used as input for the XCA, along with XRD data from a high-throughput crystallization screen.
    \textbf{e,} The experimental XRD patterns from this organic polymorph screen are low symmetry in comparison the inorganic phases studied here and they are also often complicated by amorphous or low-crystallinity impurities.
    \textbf{c,} A combinatorial materials library comprising a complete ternary system and binary sub-systems prepared in a single experiment by multilayer wedge-type nanoscale film deposition and annealing.
    \textbf{f,} Thin film XRD patterns from the library suffer from preferred orientation, phase mixing, peak shifts according to Vegard’s law, and variable noise from oxide and library edge effects. 
  }
  \label{fig3}
\end{figure}

We successfully applied XCA to solve three separate materials challenges: detecting subtle symmetry transitions in an inorganic ferroelectric, discovering organic polymorphs predicted \emph{a priori} by computation, and mapping the phase space of a metal alloy system.
These three challenges span a range of technical problems, with varying data quality and resolution.
In the first example, classification across the phase transitions of BaTiO$_3$, a canonical ferroelectric (Fig.~\ref{fig3}a), is not possible using traditional methods without expert intervention (Fig.~\ref{fig3}b)\cite{Page_2010}, but we achieved this here using the XCA.
High-throughput searches for new organic pharmaceuticals and other functional organic materials can be informed by CSP algorithms that predict energetically preferred crystal structures of a candidate molecule (Fig.~\ref{fig3}c)\cite{Collins_2017, Pulido_2017}, but XRD patterns for such materials often include a disordered background from amorphous or low-crystallinity impurities (Fig.~\ref{fig3}d); also, the CSP-derived patterns may not agree precisely with experiment for the reasons outlined above.
In our second example, we applied the XCA to adamantane-1,3,5,7-tetracarboxylic acid (ADTA), which forms a range of hydrogen bonded nets\cite{Ermer_1988, Cui_2019}.
Out third example was phase mapping of a complete ternary alloy system, Ni-Co-Al, which requires high-throughput characterization following combinatorial synthesis (Fig.~\ref{fig3}e)\cite{Ludwig_2019}. XRD patterns of thin-film samples suffer from significant and varying texturing, as well as peak shifting from the expected positions induced by strain, sample offset, and compositional variation according to Vegard’s law (Fig.~\ref{fig3}f).
The systems are also different in their data quality, as the BaTiO$_3$ data was collected using synchrotron radiation, and the others with distinct in-house powder diffraction configurations.  As such, each of these three challenges is technically distinct, but united by a need for rapid but high-quality analysis of large amounts of XRD data produced by high-throughput experiments.

\subsection{Detecting subtle phase transitions in BaTiO$_3$}
\begin{figure}
    \centering
    \includegraphics[width=\textwidth]{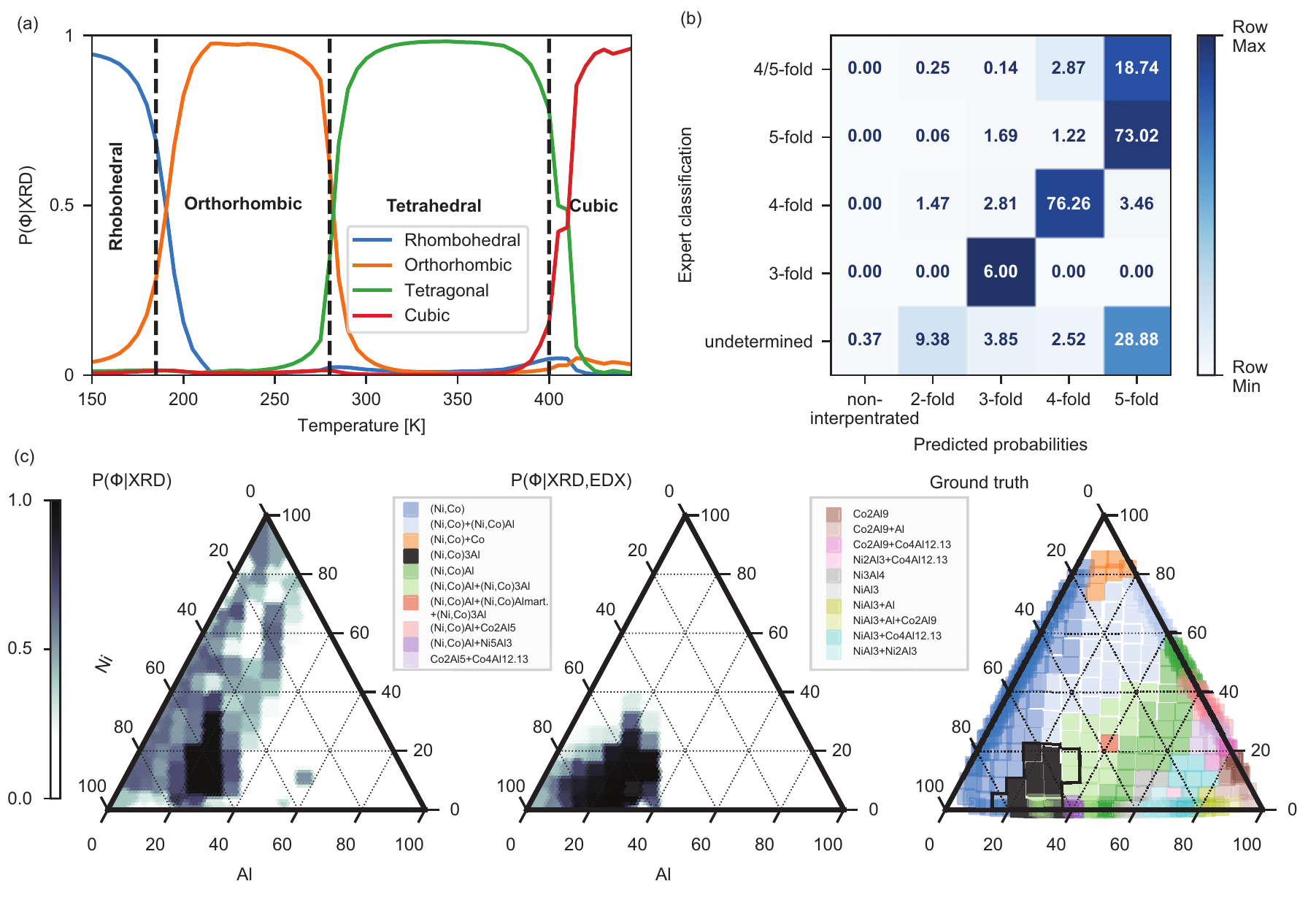}
    \caption{\textbf{Autonomous XRD analysis results from XCA.}
      \textbf{a,} XCA rapidly produces a probabilistic temperature-dependent phase mapping of BaTiO$_3$ that is more accurate than current refinement techniques. Dotted lines show the expected transition temperatures, and each colored line corresponds to the probability of a given phase existing.
      \textbf{b,} Confusion matrix showing the sum of predicted phase probabilities for each expert-classified phase of ADTA.
      \textbf{c,} Phase mapping for cubic Ni$_3$Al compared with the ground truth phase diagram with a black line outlining the probability region of $P(\phi_i |\mathrm{XRD,EDX})>=0.85$ (right). The XRD-based probability (left) captures the uncertainty associated with classifying this phase against another cubic phase with similar peak positions. The joint probability (centre) reduces the uncertainty by conflating prior information from composition. 
        }
    \label{fig4}
\end{figure}

We first tested XCA with a temperature-dependent XRD experiment across a temperature range of 150\,K to 450\,K covering four phases of BaTiO$_3$: rhombahedral (R3m), orthorhombic (Amm2), tetrahedral (P4mm), and cubic (Pm$\bar{3}$m).
BaTiO$_3$-based materials are a platform for probing the mechanisms of ferroelectric materials because their phase transitions occur at relatively low temperature\cite{Page_2010, Wegner_2020}.
Cooling from the paraelectric cubic phase induces three phase transitions that involve polarized displacements of the Ti ion along unique axes (Fig.~\ref{fig3}a), resulting in nonobvious XRD peak splitting and shifts (Fig.~\ref{fig3}b).
From four refined initial phases, XCA outputs smoothly varying probabilities across each transition temperature, and successfully identifies the mixed-phase transitions in the dataset (Fig.~\ref{fig4}a).
Almost all (56 of 60) of the classifications match the expectation, the only differences being accounted for by a temperature-lag between the measured and actual sample temperature during ramping.
The smooth variations and automatic classification represent a major improvement on current Rietveld refinement procedures, which cannot automatically identify the phases and often require additional expert scrutiny (Fig. S2).

\subsection{Searching for predicted phases in a CSP database}
We next used XCA to search for predicted organic polymorphs in a CSP structural database.
Polymorphism in organic crystals is important because different polymorphs of active pharmaceuticals, electronic molecules, and porous molecules can exhibit profoundly different physicochemical or physisorption properties\cite{Bernstein_2010}.
In our previous study\cite{Cui_2019}, high-throughput crystallization screening and XRD for a tetrahedral molecule, ADTA, generated 228 XRD patterns of varying quality. ADTA typically crystallizes to form a hydrogen-bonded network with a diamondoid topology, but CSP predicted\cite{Cui_2019} that polymorphism was likely because of the relatively small calculated energy gaps between the 5-, 4-, 3-, and 2-fold interpenetrated structures (Fig.~\ref{fig3}c).
To find these phases in our earlier study\cite{Cui_2019}, the XRD patterns were searched iteratively by eye using the CSP dataset as a structural guide.
This manual analysis of 228 experimental XRD patterns against the five lowest lattice energy CSP-derived patterns for the 5-, 4-, 3-, 2-, and 0-fold interpenetrated structures (labelled in Fig.~\ref{fig3}c) required weeks of effort, yielding a labeled test set of 187 labeled patterns, and 41 patterns which could not be labeled.
Starting from the same CSP dataset, and using the five labelled low-energy phases as inputs, XCA classified the pure phases and phase mixtures for which an expert classification was available with an experimental test accuracy of 0.952 in just a single day.
The cosine similarity, which is a measure of alignment between the output probability the ground truth and the F1-score---a metric accounting for the effects of class imbalance---both reflect this accuracy (0.941 and 0.946, respectively). This demonstration of using accurate AI for organic crystal XRD classification is unique because of these samples' lower symmetry and amorphous backgrounds (Fig. 4b).

\par

In the previously reported XRD dataset,\cite{Cui_2019} a precise match with the low-energy 2-fold phase from the CSP could not be found, but there were 41 experimental patterns that could not be characterized by the experts in reference to the CSP data. This was because variations in peak position and intensity provided insufficient information for confident classification.
As shown in Figure~\ref{fig4}b, 11 of these 41 patterns were classified by XCA as the same phase.
Of these 11 samples, 4 were able to be produced as single crystals for more extensive XRD experiments. These structures were confirmed to be the 2-fold interpenetrated phase\cite{Cui_2019}.
XCA was therefore able to correctly suggest the predicted, elusive 2-fold phase in the experimental XRD dataset, which could not be confidently classified by the research team.
This result suggests wider opportunities for the discovery of low-energy polymorphs that are predicted by CSP but where the experimental data do not agree exactly with the computational predictions; for example, because of solvent inclusion, in the case of porous materials\cite{Slater_2017}, or where the CSP does not fully capture molecular flexibility\cite{Cui_2020}.

\subsection{Using XCA to assist in phase mapping}
Lastly, we used XCA for the phase mapping of a complete ternary inorganic system, Ni-Co-Al, where composition-structure-property relationships were previously identified across 21 phase regions (Fig.\ref{fig4}c), see Figs. S5 to S24 for representative XRD patterns)\cite{Decker_2017}.
Phases in Ni-Co-Al are of interest for different applications such as superalloys and ferromagnetic shape memory applications\cite{Decker_2017,Naujoks_2016}.
Identification of the compositional existence ranges of the phases and phase mixtures requires extensive analytical effort but is critical for these materials.
Here, the XCA output probability was conflated with an independent probability, based on chemical composition from EDX,$P(\phi_i | \mathrm{EDX})$, to yield a joint probability, $P(\phi_i |\mathrm{XRD,EDX})$.

\par

When testing using the 12 phases found in the experiment, this $P(\phi_i |\mathrm{XRD,EDX})$ approached the ground truth, with most misclassifications still assigning high---but not the highest---probability to the existence of phases in a sample for both pure regions and mixtures (\emph{SI}).
To demonstrate robustness of XCA when the existing phases are unknown, we tested XCA on all unique and experimentally accessible structures of all single element, binary and ternary combinations of Ni-Co-Al in the ICSD46 (31 phases, Table S4).
Since nearly two thirds of these phases do not exist in the experiment, this approach under-performs (cosine similarity = 0.735, accuracy = 0.763, F1-score = 0.788); nonetheless, $>90\%$ of the classifications contain the correct phase in the top three probabilities. This behaviour is similar to the pattern matching approaches that propose plausible phases, but here it is effective with non-ideal thin-films and phase mixtures.
A task that previously took weeks to months of manual effort is now accelerated to take place within hours of computer time.

\par

Where XCA and the expert disagree, additional information helps to make the correct classification.
An example is shown by a low symmetry phase (NiAl$_3$) being fully textured along a specific axis, such that its pattern is commensurate with NiAl (Fig. S25): values of $P(\phi_i |\mathrm{XRD})$ and  $P(\phi_i |\mathrm{XRD,EDX})$ align with expert opinion, but are not informed by the literature or predicted phase stability.
Since  $P(\phi_i |\mathrm{EDX})$ only captures the average sample composition, both $P(\phi_i |\mathrm{XRD,EDX})$ and  $P(\phi_i |\mathrm{XRD})$ should be considered in tandem for a full phase map.
As an example, Figure~\ref{fig4}c compares the outputs for a representative phase (Ni$_3$Al) in the ternary composition space: while  $P(\phi_i |\mathrm{XRD})$ extends to encompass degeneracy (Fig.~\ref{fig1}a) and mixed phase regions, the  $P(\phi_i |\mathrm{XRD,EDX})$  is confined by the expected composition of pure Ni$_3$Al.
As for BaTiO$_3$, this produces a probabilistic solution to instantiate a precise refinement, emphasizing XCA-researcher collaboration.

\par

\section{Discussion}
The strength of XCA stems from its combination of a probabilistic model for addressing uncertainty and use of physically relevant synthetic datasets, thus allowing for applications across physics, chemistry, and biology.
Compared to cutting edge approaches, XCA is more accurate across materials systems
We compared XCA’s performance directly against the AutoXRD approach developed by Oviedo et al\cite{Oviedo_2019} considering all combinations of dataset synthesis and modelling (Fig.~\ref{benchmarking}). Details of this experiment can be found in the supplementary information.
For inorganic systems, this performance stems from the physically accurate training data, learner ensembling, and ability to incorporate additional probability.
In the case of organic polymorphs, the data production pipeline is most important because there is less XRD degeneracy between phases.
\par

\begin{figure}
  \centering
  \includegraphics{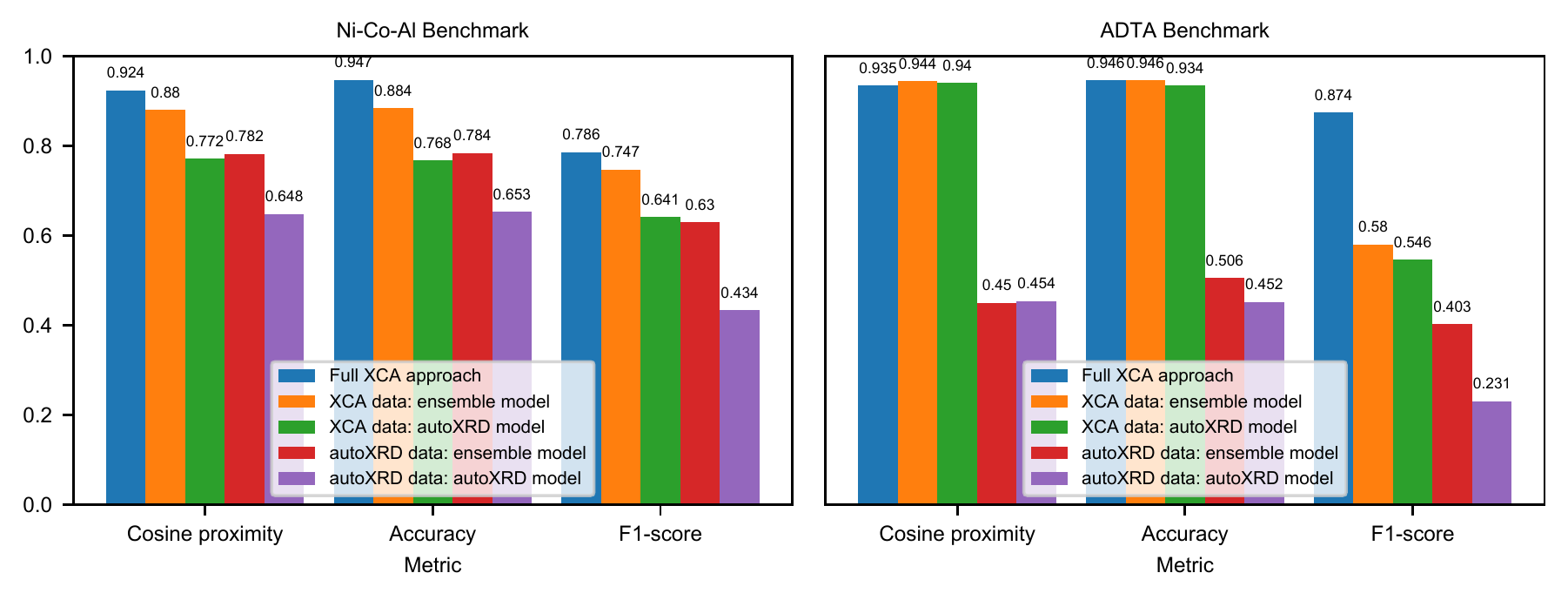}
  \caption{\textbf{Comparing different approaches to building a synthetic dataset and classifier.}
    The cosine proximity, accuracy, and F1-score (macro) are shown for (left) Ni-Co-Al and (right) ADTA pure phases using the full XCA approach, the XCA dataset with an ensemble of AutoXRD classifiers, the XCA dataset with a single AutoXRD classifier, the AutoXRD dataset with an ensemble of AutoXRD classifiers, and the AutoXRD dataset with a single AutoXRD classifer. The ensemble classification includes a joint probability distribution for the alloy system.}
  \label{benchmarking}
\end{figure}

\par

The model comparison highlights the utility of different features of the XCA.
When considering just the pure phases in the alloy dataset, the importance of ensembling and a physically accurate dataset is clear.
The information scarcity is also on display: the lower variance offered by a smaller model for each agent in the ensemble prevents overfitting and allows for marginal gains in accuracy.
The ensembling outperforms other models, especially when combined with an independent probability.
The XCA  approach is more accurate in the limiting case of textured phases producing degenerate patterns that we encounter here. In the case of the organic ADTA polymorphs---where the problem of degenerate solutions is less prevalent and there was no secondary probability distribution---there is less sensitivity to the architecture used.
Here, it is clear that the data production pipeline is most impactful. This is unsurprising, as the AutoXRD approach was designed against high symmetry perovskite structures that produce fewer peaks in the powder patterns, and organic systems often tend to crystallize in lower symmetry.
The relatively lower values for F1-score in most ADTA tests is a result of a class imbalance and the macro-averaging approach that is appropriate for pure phases.
Overall, the strength of the XCA  stems from its combination of a probabilistic model in the case of an uncertain problem, and using fully physically relevant synthetic datasets, thus allowing for applications across domains in the physical sciences.

\par

Other approaches have attempted to use synthetic XRD data to train models\cite{Lee_2020,Park_2017,Chen_2020}. Since these methods are proprietary, they cannot be compared directly; however, the study by Lee et al\cite{Lee_2020} purposefully avoided textured data, which is imperative to thin film diffraction.
Recent work\cite{Aguiar_2019} incorporating constraints into variational autoencoders has been used to solve a phase diagram problem working from a pattern synthesis similar to that used by Oviedo et al.\cite{Oviedo_2019}
This approach to demixing is a promising unsupervised approach for high symmetry materials, when there is no prior knowledge of phases and texturing is not a dominant challenge.
The development of XCA builds on insights from AI, materials science, and crystallography, and is shown effective across organic and inorganic materials systems facing a wide array of experimental complexity. 

\par

A crucial component of XCA is the ensemble of learners, and their respective uncertainty.
This uncertainty is necessary in cases where a single pattern has multiple plausible structural solutions.
In the Figures~S32-S36 and Figures~S38-S40, we explore how the Shannon entropy (a measure of uncertainty) of the posterior distribution from a single learner and ensemble of learners is effected when the synthetic dataset doesn’t capture the diversity of patterns produced by the experiment.
Critically, uncertainty is also a requirement when the training data is perfect.
As is the case with BaTiO$_3$, all four phases can be fit to every pattern with a Rietveld refinement (Fig.~S2), so even an expert should express some scrutiny in how this decision is made.
The ensembles in XCA accomplish this while still producing the correct classifications (Fig.~S37).
The problem is more complicated for the alloy system, since real uncertainty can arise from homometrics and inadequacies in the training data; however, the XCA methodology produces a rich phase map of uncertainty to accompany the existence phase map (Fig.~S41).

\par

Our methodology could be extended to any 1D response function in high-throughput materials research that requires classification and can be simulated at low cost (pair distribution function, X-ray photoelectron spectra, X-ray absorption near edge spectra, photoluminescence spectra, nuclear magnetic resonance spectra, mass spectra, etc.).
Moreover, when XCA encounters unseen phases, it will tend to broaden its output probability distribution and maximize information entropy.
To enhance this feature, future developments should diversify the architecture of individual learners and their data exposure.
To enable materials discovery in the absence of predicted phases, XCA should be extended to pair with unsupervised methods that are not necessarily conditioned on a prior. 
The collaboration between a federation of agents, including XCA and other agents fit to task, will allow for advanced materials characterization to be incorporated into adaptive learning approaches for autonomous, data-guided experimentation.

\par

In conclusion, we present an autonomous companion agent for the rapid, accurate classification of XRD datasets that is effective across materials domains, requires no labelling of experimental data, and is robust despite varying degrees of texture, peak shifting, peak broadening, phase mixing, and amorphous disorder.
The agent was designed as a probabilistic approach for challenges with substantial uncertainty.
It outputs phase maps over compositional space and discrete probability distributions per sample.
It avoids the combinatorial explosion over mixtures by probabilistically learning about pure phase existence.
The success of this approach is underpinned by ensembling ML models and the direct use of expert insight in the dataset development.
As such, it can be extended to any analysis method where a rapid, accurate simulation is available.
The XCA takes less than a day to train and enables real-time analysis during XRD measurements.
It scales effectively for more data intensive challenges involving larger multidimensional search spaces, such as developing high entropy alloys\cite{Miracle_2017} and complex solid solution electrocatalysts\cite{Loffler_2019}. The innovation is directly applicable to inverse design approaches\cite{Sanchez-Lengeling_2018}, new robotic discovery systems\cite{Granda_2018,MacLeod_2020,Li_2020,Collins_2017}, and can be immediately considered for other forms of characterization such as spectroscopy and the pair distribution function.

\section{Acknowledgements}
We acknowledge financial support from the Engineering and Physical Sciences Research Council (EPSRC) (grant number EP/N004884/1) (P.M.M., M.A.L., A.I.C.), BNL Laboratory Directed Research and Development (LDRD) projects 20-032 “Accelerating materials discovery with total scattering via machine learning” (P.M.M., D.O.), the Leverhulme Trust via the Leverhulme Research Centre for Functional Materials Design (P.C., A.I.C.), and German Research Foundation (DFG) as part of the Collaborative Research Centre TRR87/3 “Pulsed high power plasmas for the synthesis of nanostructured functional layers” (SFB-TR 87), project C2 (L.B., Y.L., A.L.). This research utilized the PDF (28-ID-1) Beamline and resources of the National Synchrotron Light Source II, a U.S. Department of Energy (DOE) Office of Science User Facility operated for the DOE Office of Science by Brookhaven National Laboratory under Contract No. DE-SC0012704.
We thank ZGH (Zentrum f\"ur Grenzflächendominierte H\"ochstleistungswerkstoffe, Ruhr-Universit\"at Bochum) and Diamond Light Source for access to beamlines I19 (MT15777) and I11 (EE17193) for XRD measurements.

\section{Author Contributions}
P.M.M., L.B. and Y.L. conceived the project. P.M.M. led the development of XCA and coordinated the research teams. L.B. contributed to development, prepared the alloy dataset, and guided the inorganic dataset synthesis. P.C. and M.L. crystallized ADTA, and measured XRD data. Y.L. advised the machine learning. D.O. measured the BaTiO$_3$ and advised the relevant studies. A.L. supervised the development and the alloy studies. A.I.C. supervised the development and organic materials studies. Data was interpreted by all authors and the manuscript was prepared by all authors.

\section{Methods}
\subsection{Synthetic dataset preparation}
Dataset preparation proceeded by collecting a set of proposed phases---from the accessible composition space in the ICSD or the local energetic minima of a CSP landscape---as crystallographic information files, and developing a large experimentally relevant set of diffraction patterns that correspond to each pure phase for the given experiment.
From the structural information (symmetry, lattice parameters, atomic positions, occupancies, and thermal displacement parameters), the multiplicities and structure factor can be calculated using the open-source computational crystallography toolbox (CCTBX)\cite{Grosse-Kunstleve_2002}.
From the experimental geometry and set-up, Lorentz polarization and an optional extinction correction can be applied.
We next applied preferred orientation randomly to each pattern. This was done by choosing a reflection plane contained in the experimental 2$\theta$ domain, and randomly varying the degree of texturing (described by the March parameter)\cite{Giacovazzo_2011}.
This drastically increases the size of the dataset and allows for a full scope of texturing to be applied to a given phase.
The peak shape is varied for each pattern using a pseudo-Voigt profile function with a random choice of mixing parameters and Caglioti parameters\cite{Giacovazzo_2011}.
A background function is randomly varied using a Laurent series with degree 6 and order -2.
The dataset generation is thus directly relevant to the experiment, encompassing the same parameters that would need to be refined during a Rietveld refinement, and depends only on a few user-defined bounds for random sampling that are inferred from experimental system: those for the background function, peak shape, and noise.
For all three materials systems, 100,000 XRD patterns were simulated for each phase.
Example dataset synthesis parameters are provided in the available code. For the BaTiO$_3$ experiment, the experimental data was high quality with good powder averaging, so the synthetic dataset included low noise, narrow peak width, moderate peak shift (even with high quality data, moderate peak shift enables robustness if the input phase has slightly incorrect dimensions), a linear background, and limited texturing.
For the ADTA experiment, the primary concerns were preferred orientation, peak shift from solvation expansion, noise from background and diffuse scatter, divergent behavior at low 2$\theta$ (accommodated by the negative Laurent series terms), and broad peak shape from diffuse scatter.
The alloy dataset synthesis was focused on texturing and noise, with a relatively constant background, and other terms equivalent to the ADTA synthesis. 

\subsection{Probabilistic model architecture}
In order to limit the overconfidence of the model, we used an ensemble of 50 shallow CNN learners. Each learner contained 3 convolutional layers (8, 8, and 4 filters, respectively) with a fixed kernel size of 5 and stride of 2, followed by a dense layer the size of the number of phases.
Dropout at a rate of 40\% was applied to the penultimate layer during training.
The final dense layers are averaged, yielding a discrete probability distribution.
The networks are trained using the Adam optimizer \cite{Kingma_2015} for 10 epochs.
A Bayesian optimization scheme was applied to optimize the hyperparameters and learner architecture. Inputs for the optimization were the number of convolutional layers, the number of initial filters, the initial stride, the initial kernel size, the number of dense nodes, and dropout rate, as well as the rate of change for each parameter of the convolutional layer between layers.
Since the validation dataset used to produce metrics for Bayesian optimization is a randomly sampled subset of the synthetic dataset, there was limited variation between training and validation. We abstained from using the test data in the Bayesian optimization scheme to avoid any ‘data leakage’.
As such XCA results are not significantly impacted by learner architecture.

\par
In the case of Ni-Co-Al, EDX measurements were available for each of the 342 samples in the experimental library. These measurements are used to construct a probability, $P$,  of phases, $\phi$, given the EDX data,
\begin{equation}
    P(\phi_i | \mathrm{EDX}) = \prod_\alpha \exp{\frac{(x_\alpha-x_{\alpha}^i)^2}{\sigma^2}},
\end{equation}
where $x_{\alpha}$ is the measured mass fraction of component $\alpha$, $x_{\alpha}^i$ is the mass fraction of component $\alpha$ in phase $i$, and $\sigma$ is set such that the full-width-at-tenth-max is 0.5. 
Treating the output from the ensemble neural net as a probability, 
$P(\phi_i|\mathrm{XRD})$
, and given that these distributions are independent, a joint probability, 
$P(\phi_i|\mathrm{XRD, EDX})$
is formed by a normalized product.
This allows the model to probabilistically differentiate between two phases appearing similar in the XRD. For example, two cubic structures of different composition that may have significant preferred orientation and strain would have similar marginal probability given the diffraction pattern, yet inclusion of the auxiliary measurement dramatically reduces the likelihood of a nonexistent phase (Fig.~\ref{fig2}).
This emulates the thought process of a metallurgist in an explicitly probabilistic way, analysing a sample by considering all of the experimental information available. This can be extended to materials systems using spectroscopic measurements where key regions of interest can be mapped to the likelihood of an intermolecular configuration.

\par

The models yield a discrete probability vector, which can than be compared against the expert classification. The test sets were not refined as quantitative mixtures, so pure classifications are converted to 1-hot vectors and multiclass classifications are converted to multi-hot vectors.
We use three metrics to measure the utility of the approach.
Since we are more interested in probabilities than absolute predictions (\emph{i.e.} argmax) and we need to understand the handling of mixtures, we first use a cosine proximity between the prediction and the ground truth as a measurement of accuracy.
In a pure system with a fully confident prediction this converges to the traditional accuracy metric, which is the fraction of maximum predicted probabilities which match the ground truth.
In the case of mixtures, this would be the fraction of the of maximum predicted probabilities that at least appear in the mixture.
Lastly, the F1-score is calculated from the global true positives, false negatives, and false positives for test sets limited to pure phases (macro-average), and aggregated from the contributions of all classes for test sets including multi-class mixtures (micro-average).

\subsection{Data availability}
The experimental datasets and code used for constructing the synthetic datasets are available as examples with the source code. Source Data for Figures 1, 3, and 4 is available with this manuscript.

\subsection{Code availability}
To facilitate the impact of this tool, the approach is kept entirely open-source under the BSD 3-clause license, and being embedded into data acquisition frameworks at central facilities (\href{https://blueskyproject.io/}{blueskyproject.io}).
Ongoing development of this tool is located at \href{https://github.com/maffettone/xca}{github.com/maffettone/xca}.
A release at the time of publication and example code for the results contained here can be found at \href{https://github.com/bnl/pub-Maffettone_2020_08}{github.com/bnl/pub-Maffettone\_2020\_08}\cite{xca_software}.
The Bayesian optimization code can be found at \href{https://github.com/maffettone/bayes_opt}{github.com/maffettone/bayes\_opt}. 

\subsection{Competing interests}
The authors declare no competing interests.


\end{document}